\documentclass[10pt]{article}

\usepackage{graphicx}

\usepackage{amsmath,amssymb,amsfonts}
 \usepackage{color}
 \usepackage{hyperref}
\usepackage{listings}
\usepackage{MnSymbol}
\usepackage{caption}

\definecolor{light-gray}{gray}{0.95}
\newcommand{\ie}{\mbox{\textit{i.e.}, }}
\newcommand{\eg}{\mbox{\textit{e.g.}, }}

\newcommand{\cws}{crowdsourcing }
\newcommand{\Cws}{Crowdsourcing }

\newcommand{\always}{\Box}
\newcommand{\sometime}{\lozenge}

\newcommand{\gwen}{\textsc{Gwendolen}}

\newcommand{\agentspeak}{AgentSpeak}
\newcommand{\jason}{{\em Jason}}
\newcommand{\tripleapl}{3APL}
\newcommand{\jadex}{Jadex}
\newcommand{\goal}{GOAL}

\DeclareRobustCommand\bfsl{\normalfont\bfseries\slshape\boldmath}

\newcommand{\mfchange}[1]{{\color{black}{#1}\color{black}}}
\newcommand{\mschange}[1]{{\color{black}{#1}\color{black}}}

  \newtheorem{definition}{Definition}

\begin{document}

\title{An Abstract Formal Basis for Digital Crowds\thanks{This work was
    partially funded by the EPSRC within the ``Verifying
    Interoperability Requirements in Pervasive Systems'' (EP/F033567)
    and the ``Reconfigurable Autonomy'' (EP/J011770) projects.}}


%\author{Marija Slavkovik \and Louise A. Dennis \and Michael Fisher}


\author{Marija Slavkovik \\
           Department of Information Science and Media Studies, \\
           University of Bergen, Norway \\
              Tel.: +47-55582377\\
              \texttt{marija.slavkovik@infomedia.uib.no}
           \and
           Louise A. Dennis and Michael Fisher \\
           Department of Computer Science, \\
           University of Liverpool, United Kingdom \\
              Tel.: +44-1517954275\\
              \texttt{l.a.Dennis@liverpool.ac.uk}
}

\date{Received: date / Accepted: date}

\maketitle

\begin{abstract} 
Crowdsourcing, together with its related approaches, has become very
popular in recent years. All crowdsourcing processes involve the
participation of a digital crowd, a large number of people that access
a single Internet platform or shared service. In this paper we explore
the possibility of applying formal methods, typically used for the
verification of software and hardware systems, in analysing the
behavior of a digital crowd. More precisely, we provide a formal
description language for specifying digital crowds. We represent
\mfchange{digital} crowds in which the agents do not directly
communicate with each other. We further show how this specification
\mfchange{can} provide the basis for sophisticated formal methods, in
particular formal verification.
\end{abstract}

\section{Introduction}
% 
%Define your pasture -- what is crowdsourcing?
%
\Cws is an umbrella term used to describe a wide range of activities
coordinated within an Internet-based platform. It's original definition is by Howe:
\begin{quote}
\emph{Simply defined, crowdsourcing represents the act of a company or institution taking a function once performed by employees and outsourcing it to an undefined (and generally large) network of people in the form of an open call. This can take the form of peer-production (when the job is performed collaboratively), but is also often undertaken by sole individuals. The crucial prerequisite is the use of the open call format and the large network of potential laborers.}~\cite{Howe:2006}
\end{quote}

\noindent \mfchange{In the most general sense,} therefore, \cws can be
defined as the business practice of a ``seeker'' entity, attempting to
out-source activities, such as solving problems, performing tasks or
raising funds, to {\em a priori} unidentified undertakers, namely the
``crowd''.  \Cws has a wide range of potential applications and is the
basis for many successful businesses; as such it has also emerged as a
new and vibrant research area~\cite{ZhaoZ:2012}.

The ubiquitous component of all the \cws processes is the \emph{crowd}
or, more specifically, the {\em digital crowd}. In English, the term
``crowd'' is used to refer to a large number of people that are in
each other's immediate presence~\cite{Lofland:1993}. To distinguish
our terminology from this we use ``digital crowd'' to refer to a large
number of agents that access a single Internet platform or shared
service, \ie a large number of agents that are in each other's
immediate e-presence. The agents can represent individual people,
organisations, other web-based services, etc.

\Cws is, foremost, a business practice and the methods of managing and
analysing it are ``practical'' rather than ``formal''.  To the best of
our knowledge, there are no formal methods   to verify various
requirements, for example, to determine when \cws is an adequate
approach to a problem, under which conditions \cws will be successful,
how effective \cws can be on a particular problem, whether a digital
crowd can be expected to solve a problem successfully, etc. Our aim is
to explore whether formal methods can be applied to improving the
understanding and \mfchange{effectiveness} of digital crowds. 
%% \\\texttt{\%Might be nice to say something more about formal
%%  methods and specification and the advantages to using them.}

\subsection{Varieties of Digital Crowds}
Originally covered by the umbrella term, ``crowdsourcing'', a range of
applications involving digital crowds   have been developed,
incorporating many different agent structures and different
interaction protocols. Some initial attempts  to develop a taxonomy for the
growing number of \cws varieties, have
been made~\cite{SchenkG:2011,SpanishPaper}. 

We consider some, but by no means all, crowdsourcing practices, from the viewpoint of the interaction between the seeker and the digital crowd and the communication among the digital crowd participants. We would like to emphasise that because of the lack of a comprehensive common taxonomy, and since the terms for different processes partially overlap among different taxonomists, the names used for
each process here may be considered imprecise by some or inadequate by others. We follow to some extent the crowdsourcing taxonomy developed  in \cite{SpanishPaper} as a recent  approach that encompasses and builds upon earlier work.

%To highlight some of the digital crowd variations of what is covered under the
%``crowdsourcing'' banner, we briefly describe five approaches, and
%give examples of existing business/charity platforms for each. (Again,
%because of the lack of a comprehensive taxonomy, the names used for
%each process may be considered imprecise or inadequate.)

\paragraph{Crowdcasting} is possibly the earliest  and best understood example of a \cws
process. Here, a seeker advertises an open challenge to the crowd and
offers a financial reward for  one solution, \eg the best or the first. Seekers can be
people, businesses, or representatives of larger institutions, but
each is represented by, and acts as, one agent. The challenges can
range from small creative tasks such as the design of a logo, to grand
challenges for solving complicated technical problems. The crowd that
responds to crowdcasting is usually comprised of people that have the
adequate skills to solve the challenge, though no certification is
required that participants possess such skills. People can solve the
tasks individually, or work in teams. However, it is not the aim of
the platform to facilitate the finding of team-mates.  

\mfchange{The} \emph{99designs} ({\small\url{99designs.co.uk}})
\mfchange{system} is a platform that enables seekers to advertise
their design needs and proposed rewards. The platform users wishing to
respond to a seeker do so by sending in their design proposals. The
seeker chooses the preferred design solution and rewards the proponent
of that solution. The respondents are typically individuals and all
communication is done via the platform. In contrast, {\em Innocentive}
({\small\url{www.innocentive.com}}) is a platform where difficult
technical challenges are posed, for which a solution may not even
exist. Here, the respondents are typically teams of people who already
know each other, with varied skill sets, that share the prize if
awarded.

\paragraph{Crowdcomputing} is a \mschange{general} term we use to name the \mschange{many} processes whereby a seeker defines an end goal that is to be accomplished by the crowd participants. Each of the participants contributes a fractional effort towards accomplishing that goal. \mschange{One of these processes is}  microtasking, a process where a vast number of very simple tasks that take little time and effort to solve are created by the seeker. The task solvers are almost exclusively individuals, and the
rewards are small. The rewards for the crowd might be monetary or an
intrinsic motivation maybe induced in the participants to inspire them
\mfchange{to contribute}, \eg altruism or entertainment.  As in
crowdcasting, the platforms do not offer \mfchange{any} means for
communication among members of the crowd. 
\mfchange{However, in contrast to} crowdcasting one task is given to
one solver, \mfchange{and} thus all solutions are paid for.  To ensure
that the solutions are of a satisfactory standard, the solvers must be
certified before they can be assigned tasks.

Representative examples of microtasking crowdcomputing platforms with financial rewards are
Amazon's \emph{Mechanical Turk} ({\small\url{www.mturk.com}}), and
CrowdFlower ({\small\url{crowdflower.com}}).
An example of a crowdcomputing process where the rewards are entertainment rather than monetary are when they are  organised as a {\em
  game with a purpose}. Here, the game is hosted by the platform, the
  participants solve the tasks by playing the game and their reward is
  the pleasure derived from playing.  In this type of crowdcomputing, pre-certification is not
  necessary. A typical example is the DNA puzzle game for
  multi-sequence alignment Phylo developed by the McGill Centre for
  Bioinformatics ({\small\url{phylo.cs.mcgill.ca}}).

Common \mfchange{to} all these processes is the fact that although the crowd participants solve a problem jointly, they do not actually collaborate, are not aware of, nor depend on, the solutions of the other crowd participants. 

\paragraph{Crowdcontent,}  as described by \cite{SpanishPaper}  includes the processes  in which a content is created, analysed or discovered by a digital crowd.  The term  ``Wisdom of the
Crowds'' or ``Collective Intelligence'', can also be considered to
include these processes.  The abilities, information and motivation of
the digital crowd's participants are combined, resulting in the
emergence of new abilities, information and intentionality. While to
some extent this is true with crowdsourcing processes such as
Mechanical Turk, there is a notable difference; in Mechanical Turk the
tasks, called ``human intelligence tasks'' require no particular
cognitive effort to perform and all tasks requested by each individual
seeker are similar. With collective intelligence, the tasks are not
given to the crowd, but the members contribute cognitively more
demanding efforts in response to what they perceive is necessary to
complete the larger goal. Furthermore, while in the crowdcomputing
processes the participants work independently, here they consider and
build upon the work of others, \mfchange{though} direct communication
among the participants is not strictly necessary.  In
\cite{SpanishPaper} the crowdcontent processes are further
differentiated into crowd production, crowd searching, and
crowdanalysing, the terms being self-explanatory.

While collective intelligence is a phenomenon that occurs
spontaneously \mfchange{in some systems}, platforms can be designed
specifically with the task of promoting such collective efforts. In
this case, the platform is the \emph{seeker}, looking for information
and abilities, as well as assigning tasks to the digital
crowd. Possibly the best example of this \cws activity is
\emph{Wikipedia} ({\small\url{www.wikipedia.org}})\footnote{Though the
  lack of unanimity from the crowd makes Wikipedia an atypical
  example. We are grateful to the anonymous reviewer who pointed this
  out.}.  Typically no material payment is associated with this
subtype of crowdcomputing process, the reward being advancement of a
common good.

\paragraph{Crowdvoting}  is a \cws process in which a platform
only promotes or sells those goods and services that are supported by
the majority of the digital crowd. In another variant, crowdvoting is
a method of eliciting an opinion only from the community that is
interested in a particular issue. An example of a crowdvoting platform
is \emph{Threadless} ({\small\url{beta.threadless.com}}). Threadless
is a platform on which the seekers are \emph{designers} who can submit
garment illustration designs, while the crowd assigns approval to the
designs that they like. Designs that achieve a certain number of votes
are printed and offered for sale on the platform, with profits being
transferred to the seekers.

An example that conceptually lies between crowdvoting and
crowdcomputing is a collaborative effort such as writing a scenario
for a movie together, as instigated by Paul
Verhoeven\footnote{\url{http://www.bbc.com/culture/story/20130807-the-public-cant-write}}
or drafting alterations to a constitution, as is the case with
Iceland\footnote{\url{http://www.wired.co.uk/news/archive/2012-10/23/iceland-crowdsourced-constitution}}.

The processes \mfchange{by} which the crowd chooses an item, such as
in \emph{Threadless} ({\small\url{beta.threadless.com}}) \mfchange{is}
termed \emph{crowdopinion} in \cite{SpanishPaper}, while the processes
\mfchange{by} which the crowd changes the good they are creating until
there are no objections, as is the case with writing a scenario or a
constitution, is termed \emph{crowd storming}. What these processes
have in common is that unlike the others mentioned earlier,
\mfchange{crowd} participants can appear to interact with each other
directly, and even collaborate directly towards the creation of the
end good. However, even in these processes, the communication is
actually executed via the hosting platform, namely the participants do
not send private messages to each other and all communication is
accessible to the whole digital crowd.

\paragraph{Crowdfunding} is different from other \cws processes, and one of the least ambiguous. 
In crowdfunding  the seeker does not out-source a task, but advertises a
fund-raising project. The digital crowd does not compete, but
participates, as they are donors that pledge their own funds and
resources towards the accomplishment of the project. A seeker may
directly advertise a project or it may do so trough a moderator. As in
crowdcasting, seekers and donors can be individuals, teams, or even
companies, but crowd members are not expected to communicate with each
other or collaborate.  The seekers can look to raise funds to finance
their business venture, micro-loans or charity offers for non-profit
organisations.

A crowdfunding platform that matches seekers looking to finance a
business ventures is \emph{Kickstarter}
({\small\url{www.kickstarter.com}}). The donors' funds are given with
no return on investment up to a certain value or with some minor goods
in return over a certain value. \emph{Kiva}\footnote{Although Kiva was created  before crowdfunding became popular.}
({\small\url{www.kiva.org}}) is a platform that manages micro-loans to
private individuals. More precisely, Kiva first establishes contact
with moderators, which are traditional loan-givers and repayment
enforcers, and a seeker who is requesting a loan from the moderators. The stories
and needs of the seekers are passed on to Kiva by the
moderators. These stories are advertised by Kiva to the donors who
then transfer money to Kiva, which distributes it to the
moderators. The money is loaned to the seekers and, when returned, is
repaid to the donors. Kiva's task is to manage the moderators to
ensure that funds end up in the hands of the seekers. Platforms that
handle traditional fund-raising projects are \emph{Global Giving}
({\small\url{www.globalgiving.org}}) for charity and \emph{Indiegogo}
({\small\url{www.indiegogo.com}}) for both charity and personal
projects.  

\paragraph{Smart Mobs} represent a hybrid between a
digital and a physical crowd~\cite{Rheingold:2003}.  Smart mobs are not a crowdsourcing process, however we do consider them here because they involve a digital crowd. 

Smart mobs
do not utilise a \cws platform exclusively for their purpose, as these
are one-off activities. A generic platform, such as an Internet forum, is used to coordinate the activities of an actual
crowd, which as a result is formed deliberately, is purposeful and is
efficient. A seeker, typically an instigator, advertises a cause and gives the specifications of the activity he or
she wants to occur. Those from the digital crowd that wish to
participate respond by following the activity specification and, as a
result, an actual physical crowd may form.

A concept related to smart mobs is {\em Internet vigilantism}. As in a
smart mob, a seeker posts a cause and a description of activity to a
generic platform while responders participate by complying. A seeker
may act as an individual in a single instance of vigilantism or as a
complex entity that frequently takes on vigilantism activities \eg
\emph{Anonymous}~\cite{Olson:2012}. Seekers in the context of
Internet vigilantism engage in a wider range of activities that may or
may not result in physical crowd formation. For example, a seeker may
describe a perceived injustice and solicit the help of the digital
crowd to seek out the perpetrators and expose their identities to the
authorities or to the public. Seekers may also advertise a smart mob
activity but with the purpose of vigilantism in the cyber space, such
as a coordinated {\em distributed denial-of-service attack} to an
Internet service.

Common to all digital crowds we consider here is that the participants in the crowd do not communicate directly and privately with each other.  In the rare instance when there is a need to communicate with another crowd participant, the participant directs this communication to the crowdsourcing platform that hosts the crowd, and it is the platform that relays the message. In the rest of this work we focus on an abstract digital crowd, not specific for any one crowdsourcing process, in which the communication between crowd and seeker, as well as among crowd participants, is executed via a mutual crowd hosting platform. 

\subsection{Formal Methods}
Formal methods represent a collection of techniques and tools for the
specification, development and verification of software and hardware
systems. These techniques are typically based on mathematical
structures and formal logic and so provide clear and unambiguous
descriptions of required behaviour. This logical basis then provides
the opportunity for sophisticated analysis tools to be utilised, for
example based on logical proof or exhaustive state-space exploration.
Verification is the process of establishing that a designed system has
its intended properties. Formal verification has become widespread,
enabling deep and (semi) automated formal analysis of systems and so
providing greater clarity concerning reliability and correctness. The
behaviour of complex systems, even in a restricted environment, can
easily become unpredictable making verifiability a very desirable
quality in such systems.

Our long-term aim is to use an automated formal verification technique
called \emph{model-checking}~\cite{Clarke00:MC} to analyse digital
crowd systems.  However, in order to achieve this a strong and
logically precise logical basis for crowd systems must be developed;
this is the issue we address here.  Model-checking has previously
been applied to the analysis of \emph{multi agent
  systems}~\cite{RaimondiL07,MCAPL_journal} so it is a plausible
technology to target for the analysis of digital crowds. A Multi Agent
System (MAS) is an artificial system containing (possibly among other
entities) a set of agents and managing the communication between
them. Although a digital crowd is a system of agents, it cannot be
considered to be straight-forward MAS in the traditional sense of this
paradigm (we elaborate on this in \mfchange{Section}~\ref{problem}).

\subsection{Multi-Agent Systems (MAS)}
The definition of what constitutes an `agent' varies among
disciplines. In computer science and artificial intelligence, an agent
is typically defined as
\begin{quote}
 ``an encapsulated computer system that is situated in some
  environment, and that is capable of flexible, autonomous action in
  that environment in order to meet its design
  objectives''~\cite{Jennings:1998}.
\end{quote}
Further, an agent is usually considered to be responsive with respect
to other agents and its environment, pro-active in its activities, and
able to interact with other agents where necessary. Analysis and
construction of MASs is central to issues such as problem solving,
coordination, cooperation and control with, and of, multiple
agents. The agent paradigm, and the ensuing methods for modelling and
representing agents and systems, are geared towards supporting these
research aims.

The \emph{specification} of a MAS means building a representation of
the agents, and possibly the enclosing environment, that constitute
the system. A common route to specifying more sophisticated agents
involves attributing mental states to them. The well known
\emph{Belief-Desire-Intention} (BDI) model~\cite{rao:95b} uses:
\begin{itemize}
\item \emph{beliefs} to represent the agent's (possibly incomplete and
  incorrect) information about itself, other agents, and its
  environment; 
\item \emph{desires} to represent the agent's long-term aims; and
\item \emph{intentions} to capture aims that are actively being
  pursued. 
\end{itemize}
The agent then \emph{deliberates} over what desires to target, how to
tackle them and what actions to take. The BDI model has been very
influential and has consequently been altered and extended with a
range of other mental states such as actions, obligations, abilities
etc, and various logics, such as epistemic, doxastic, deontic and
temporal logics, have been developed and successfully applied to
represent these states.

Unlike typical MASs, digital crowds are not \emph{deliberate}
systems. In a MAS each agent is an individual that is fully identified
by her mental state. Illustratively, the agent is the entity of
interest and we specify the environment and the other agents,
including the MAS, as seen through her ``mental eye''.  Even when a MAS of several agents is specified, all agents in the system are specified individually one by one by representing their mental states. 

In a digital
crowd,  no part of the participants' mental states may be made available to other agents as part of the process in which they are involved. Therefore we cannot represent the crowd participants one by one, as none of their mental states may be known.

What is known of the digital crowd participants is the messages they exchange within the crowd. It is the information exchanged between the agents that are the focus of our interest, because this is where interesting phenomena can
emerge. We propose that such exchanges are what need to be specified in order to
specify  a digital crowd.

%aware
%of both its environment and of other agents, is represented as such,
%and adjusts her behaviour in that regard when pursuing her goals.  In a
%MAS, each individual agent matters to the system in the sense that if
%an agent leaves or joins the system, this event is reflected on the
%rest of the system.  In a \emph{digital crowd} the agents constituting
%the crowd are not necessarily coordinating or cooperating with each
%other; they may not even be aware that they are part of a crowd at
%all. Individual agents, their presence, their goals, their individual
%mind states, are obscured, and these agents are Despite the agents in the crowd being
%autonomous and pursuing their own interests, the crowd functions as a
%whole and is interacted with as a whole by outside agents, such as the
%seekers in a crowdsourcing process.
 
%  Despite the
%differences between a ``traditional" MAS and a digital crowd, and
%between a \cws and ``traditional" agent interaction, conceptualising a
%digital crowd as a MAS can potentially make \emph{formal methods} for
%engineering and analysis of MAS applicable to \cws processes and
%digital crowds.
 
Agent-oriented software
engineering~\cite{Jennings00,CiancariniW:2001} is a research area
specifically concerned with developing methods for engineering
applications conceptualised as MASs~\cite{Mascardi:2004}. Within
agent-oriented software engineering, logic-based formal methods are
used to specify systems, program systems, and verify
systems~\cite{DixFisher13:chap}. The agent specification languages
that drive current agent model-checking systems are inadequate for
specifying the structural complexity of digital crowds (see our
discussion in \mfchange{Section}~\ref{problem}). In this paper, we
develop an improved logic-based agent specification language that
overcomes the shortcomings.  Communication among agents in a MAS is
considered to be an act and implemented as such, see \eg
\cite{deBoer:2003,McBurney:2005,Wooldridge:2000}. Here we are not
concerned with how the agents exchange messages, but with the content
of such messages as a special, public part of the mental state of the
agent in a social setting.

%% \emph{Model-checking} is a variety of formal
%% verification in which all possible executions of a system can be
%% examined automatically based on a model of the
%% system~\citep{Clarke00:MC}. Given a model of a system $\mathcal{M}$,
%% represented in some formal language, and a requirement, or desirable
%% property, specified as a formula in a logical specification language
%% $\varphi$, it can be checked whether the requirement is satisfied by
%% the model, namely if $\mathcal{M} \models \varphi$ holds. If the
%% requirement is specified in a temporal logic~\citep{Practical11}, it
%% can be checked whether that requirement is met by the system at some
%% specific moment, at some point in the future, or at all points in the
%% future (or some combination of these). Model-checking has indeed been
%% extended to MAS. However, though methods have been developed that are
%% scaleable~\citep{RaimondiL07} And address sophisticated BDI-like
%% agents~\citep{MCAPL_journal}, there is no approach that yet achieves
%% both.

%Maybe here Discuss relationship among agent logic and agent
%programing etc. Louise AJPFL stuff.

The paper is structured as follows. To distinguish the requirements
for 
\mfchange{an improved agent specification language for digital crowds}, 
we first discuss in more detail the difference between the
``typical" MAS and digital crowds and the shortcomings of existing
agent logics-based languages and logics in Section~\ref{problem}.  In
Section~\ref{extension} we propose an extension to the BDI agent
\mfchange{paradigm, and in} Section~\ref{s:specification} we propose an agent
specification language.  In Section~\ref{examples} we present examples
of verification routes for crowdsourcing using our specification.  In
Section~\ref{summary} we discuss conclusions and future work.

\section{Problem Analysis}\label{problem}
 
% \%\texttt{ I think we might be better off without this list of 
%% different digital crowds. I have commented out that text but 
%% you can put it back if you disagree}
%%
%We provide an overview of digital crowd structures,
%particularly from an \emph{agent} perspective; this will motivate the
%development of our logical framework in later sections.
%
Within crowdsourcing platforms, a digital crowd is typically considered
as a single entity (not an agent structure) whose intentions, beliefs
or goals cannot be known, but whose incentives can be foreseen. So,
while being fully aware that there could be vast numbers of agents
accessing the platform, and that the content of the communication is visible to all these agents, the \emph{seekers} communicate with the digital
crowd as they would with one particular agent.
This one agent is an
embodiment of the crowd, not its leader, but rather a representative agent;
a typical agent in the crowd that embodies all the properties that the seeker finds desirable in the  agent(s) he or she would ideally task with their problem.

Another common facet of crowdsourcing is that agents within the
digital crowds are not expected to interact with each other privately,
or even directly communicate with each other (some platforms that host
crowdsourcing processes do not even provide the means for such
interactions). Nevertheless, the participants are encouraged to
communicate with other agents outside the digital crowd by sharing
information involving the crowdsourcing process, and this type of
communication is facilitated by the platform. Consider for example,
the \mfchange{``share on social network''} and
\mfchange{``email to a friend''} buttons embedded in many such
platforms.

Although an agent may not be allowed to join a digital crowd if
\mfchange{it lacks} certain skills or characteristics, within the
digital crowd no informational or motivational attitudes are
shared. Even in the case of smart mobs, the motivations and beliefs
driving each agent to participate in the mob may differ. If enough
agents happen to have similar states of mind, a mob will
occur. Excluding collaboration and the sharing of mental states from
the agent structure makes a digital crowd different from a
``traditional'' MAS.

Horling {\em et al.} \cite{HorlingL:2004} give an extensive overview
of multi-agent structures such as coalitions, teams, societies etc.
We provide a brief overview to compare these structures with digital
crowds.  We first distinguish between a \emph{shared attitude} and a
\emph{we-attitude}. In the literature the we-attitude is called a {\em
  joint attitude}, but we \mfchange{here} use we-attitudes to
distinguish better from the shared \mfchange{attitudes}. The
difference is subtle: a shared attitude is an attitude that conditions
participation in a structure, \eg if your goal is to achieve
$\varphi$, then you are part of the agent structure. Illustratively,
since your colleague John wants to play football, he is part of the
university football team. In contrast, a we-attitude is an attitude
that characterises participation in a structure, \eg we (as an
organisation) have the goal to achieve $\varphi$, so if you are part
of the structure then your goal is $\varphi$ as well. Illustratively,
we are married therefore we-intend to abstain from intimate relations
with other agents.
 
A {\em coalition} is an agent structure in which each agent is
committed to pursuing a we-goal.  An agent {\em institution}, or
society, is a structure in which members have certain we-beliefs
and we-obligations of compliance with certain norms.  This does not
mean that if a norm is violated the entire institution is subject to
punitive measures, rather that if an agent is a member of the
institution, then he or she is subject to the institutional norms and
sanctions.  Collaboration is not essential, either in coalitions or in
institutions.  A {\em team} is a structure of agents that collaborate,
by assuming different roles, towards accomplishing a shared goal.

Digital crowds, and crowds in general, exhibit a dual nature: both that of a
single agent and a structure of agents.  On one hand, the whole crowd
can be viewed as one rational agent. The crowd can be given a task,
can be considered accountable for transgressions \eg in the case of
{\em smart mobs}~\cite{Rheingold:2003}, where the crowd exhibits
behaviour emergence and abilities not pertinent to any one participant.
On the other hand, the participants in the crowd are individual
rational agents, in the sense that they are fully autonomous,
undertaking only those tasks they choose to~\cite{ZhaoZ:2012}.
 
Comparing how agent structures are specified, we can observe that in
coalitions, teams, even institutions, the agents either share mental
attitudes or adopt we-attitudes. Consequently, it is possible to
represent the agent structure bottom up, by representing all the
constituent agents' shared and/or we-attitudes.  However, neither
crowds nor crowd participants can be modelled in this manner since, in
a digital crowd, agents neither share attitudes nor adopt we-attitudes. Therefore, we cannot obtain a digital crowd specification
using the bottom up approach of specifying each individual agent of
the crowd. Furthermore, specifying each crowd agent is both
unnecessary and difficult, as the number of agents in the crowd is
potentially vast and constantly changing.
 
To specify a crowd, we must encapsulate (hide) the mental states of the agent
in the representation. This allows us to represent agent structures,
such as digital crowds, as single agents within larger structures,
such as platforms. We can represent each agent with the attitudes it
decides to reveal to its environment or the structures with the
attitudes that all its members have in common.
%The current agent structures are constructed bottom up: first agents
%are specified, by specifying their mental states,   then the agents
%that adopts joint mental attitudes get to constitute an agent
%structure.   
By introducing encapsulation, we effectively construct structures top
down: first the agent structure is represented, then agents are
distinguished by adding more detailed information about their mental
states.

Agent specification languages do not represent
communication. Collaboration in agent structures is
\mfchange{formally} represented as a we-mental attitude. Eventually
the act, not the content, of communicating is represented as an
action.
%% Agent based programming languages do have explicit mechanisms for
%% sending and receiving messages, however an agent does not reason
%% with the content of these messages. These messages are limited in
%% content, expressing simple singles such as stop and go. 
How can we specify an agent in an agent structure without specifying
his/her mental attitudes? 

%% \texttt{I liked reviewer 3's comment that the beauty of the BDI is
%% that people can understand it by applying it to themselves and the
%% same is expected for our model. This is what I would like to
%% achieve bellow. Could it be expressed better?}

Consider an imageboard, an Internet forum where members can post
images available to be viewed by other members or visitors without
restriction. The forum is a context \mfchange{that} contains a crowd
of unspecified agents that pop in and out of existence. This forum
context exhibits the characteristics of an agent and, clearly, the
members and visitors are also agents. How can we represent this system
of agents without peering into the mental states of the agents
involved?  Mental states aim to capture the behavior of an
agent. Publicly accessible messages passed between an agent and a
structure, \eg crowdsourcing platform, can be used to capture the
behavior of a complex system of agents.  Such messages can be either
sent \emph{to} the agent or sent \emph{by} the agent. When accessing a
forum either as an administrator, visitor or member, the messages
posted on the forum are what can be seen and are what is used to make
deductions about the system as a whole and individual agents within
it. We need to be able to express not only the fact that a message has
occurred but also what the content of the message was, so that we are
able to represent the information that agents wish to disclose to
their environment, to give incentives to digital crowds, and allow
agents to reason with them.

%% \texttt{I added the paragraph bellow because due to Reviewer 2 I
%% see how ACL can be confused with what we do. Should we say more?
%% Can we say it better?}

A clear distinction between agent specification and the field of agent
communication is needed. Agent communication is a vast area of research within engineering multi agent systems, with considerable work having taken place on developing communication protocols   \cite{deBoer:2003,McBurney:2005,PittM:1999,Wooldridge:2000}.
Agent communication
languages~\cite{Singh2003}, such as the FIPA-ACL proposed by FIPA
({\small\url{http://www.fipa.org}}), and increasingly captured as OWL
ontologies~\cite{FIPA-ACL-OWL2}, are languages designed to be used by
agents to communicate among themselves within a MAS. 

Within the research in agent communications, the communication between
agents is represented as an act, in the sense that for example, an
autonomous vacuum cleaner is capable of vacuuming dust and
communicating its location to its docking station. While consideration
is given to the allowed content of the messages exchanged, the agent
does not internalise, \ie represent as a mental state, the fact that a
message happened. This is typical of all acts when MAS are
implemented. For example, the fact that dirt is collected by the
vacuum cleaner is not internalised after the cleaner vacuums. The
agent can detect a change in his environment (its dirt container is
now non empty) as an effect of having successfully executed the action
of vacuuming, but no information state is added that represents
\mfchange{the fact} that vacuuming has taken place. The autonomous
vacuum cleaner will receive a message with content ``stop" from the
owner, and as a result of that message it will stop its actions, but
the fact that a message with content stop has been received is not
part of the informational states of the cleaner.

We are here not concerned with communication protocols or the implementation of the  communication acts among agents. Instead, we are exclusively concerned with representing, as information states, the information that a message has been sent, or received, and  the content of that message. So we represent the intermediary step, the addition \emph{to} the information state of the agent of the fact that communication happened.  This happens {\em after} the communication {\em act} was executed.

We have to further make clear the difference between a specification and communication language. A specification language is designed to describe the overall behaviour
of a system, such as \eg a multi agent system. We specify what is said
to, and by, an agent, but this may not be the language that the agent
actually uses to execute communication. Consider an illustrative
example. An example of agents communicating using English is when John
says to Mary: ``I will go to play football." An example of English
used as a specification language is when we as the designers of this
example describe the John-Mary interaction: John has informed Mary of
his intention to engage in the football play activity. English, as
logic, can be used for both communication and specification, but we
are here concerned only about the specification aspect. For the
challenges regarding agent communication languages in social settings,
which do have an overlap with the challenges for specification
languages for systems of agents, one can see for example
\cite{Singh2003}.
\medskip

%\noindent We next consider encapsulation and specification of message passing.
 
%is limited to sWhat can be represented is the interaction among seeker and digital crowd, the content of the advertised information by the seekers and the responses of 
% encapsulate the crowd in a single agent  that only has the comment traits. do not need to represent mental states...need to reason about communications
% 
% all agent programming languages allow for communications but the content is not included in the reasoning. 
% agent logics, such as dynamic epistemic logic , go into too much detail, to obtain the state od mind of an organization, the mental states of each member are an open book, these languages build structures bottom up, we want to build structures top down. 

\section{From Agent to Agent Structure} \label{extension}
% A predominant view on rational agency is embodied by the prevailing
% Belief-Desire-Intention (BDI) approach \citep{Rao1991}, but also in
% other works of agent theory \cite{vanLinder:1998}, where many logics
% have been developed.  
% \texttt{Here reviewer 1 says that we give a different definition 
% for BDI than the introduction, and we should not repeat. Do you agree?}
%
A \emph{rational} agent takes action in order to achieve its goals
based on the beliefs it holds about its environment and other
agents~\cite{WooldridgeJ:1995}. A rational agent is typically
described by representing its dynamic, informational and motivation
aspects, \ie its mental attitudes. This often conforms to the BDI
model~\cite{rao:95b,Rao1991}\mfchange{;} recall that``BDI'' denotes
\emph{Beliefs}, \emph{Desires}, and \emph{Intentions}.
%Beliefs represent the information the agent has about the world, desires represent the long-term goals of the agent, while intentionsare goals that the agent is actively pursuing. 
There are \emph{many}
different agent programming languages and agent platforms based, at
least in part, on the BDI approach. Particular languages developed for
programming \emph{rational} agents in a BDI-like way include
\agentspeak~\cite{rao96:agentspeak}, \jason~\cite{MAPlpaX:Bordini},
\tripleapl~\cite{3APL98}, \jadex~\cite{PokahrBL05},
\goal~\cite{boer07}, \textsc{SAAPL}~\cite{winikoff07:_implem}, and
\gwen~\cite{dennis08}.

\begin{figure}
 \centering
       \includegraphics[width=0.5\textwidth]{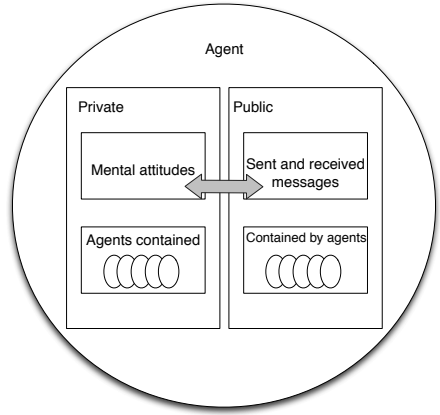}  
                \caption{Structure of an extended agent.}\label{fig:agt}
                \vspace{-0.6cm}
\end{figure}

In Fig.~\ref{fig:agt} we present a diagram of an extended agent that
will also be used to specify agent structures. The mental attitudes of
the agent are encapsulated in the \emph{private} section of the agent,
while the \emph{public} section contains the information that has been
sent to the agent, and the information sent by the agent to others.
In addition to the public and private sections, there is reference to
the other agents (specified in the same manner) that are contained in
(respectively, contain) this agent. So, if a single agent is being
specified, the list of agents contained within it is empty.
 
We can see, from Fig.~\ref{fig:agt} that there is a connection between
the private and public sections. This is relevant once the agent
decides to internalize some of the information it receives into its
mental states, or when it wants to share some aspect of its mental
states with its environment. 
%% We first discuss the encapsulation of
%% agent structures and then develop the public agent section further.
        
\subsection{Encapsulation}
An agent is typically considered to be an autonomous software or
hardware entity capable of perceiving its environment, pursuing its
own goals and interacting with the environment and with other
agents. Initially, it might appear that the agent paradigm is not
appropriate for modelling digital crowd participants, since it applies
to software and hardware entities, while digital crowds are comprised
almost exclusively of people. However, people do not directly
participate in the digital crowd --- their participation is almost
exclusively by medium of a user account, effectively their software
avatar. We can build a representation of the avatar and extend it with
a representation of the expected behaviour of the agent.
 
The formal description of flexible agent structures was developed
in~\cite{Fisher99} in the context of executable agent
specifications. There, to the formal specification of the agent's
mental attitudes, two additional sets are added: {\em content} and
{\em context}. These correspond to the two sets given in
Fig.~\ref{fig:agt}, and allow groups, teams or organizations of agents
to be dually treated as individual agents.  For example, as the
environment is described as a set of agents within the context then,
if a message is to be sent into the environment, it is just broadcast
to the context agents. Crucially, there is no distinction between an
agent that contains others and one that does not; effectively,
organisations such as crowds can be themselves viewed as agents.

\begin{figure}[h!]
        \centering
       \includegraphics[width=\textwidth]{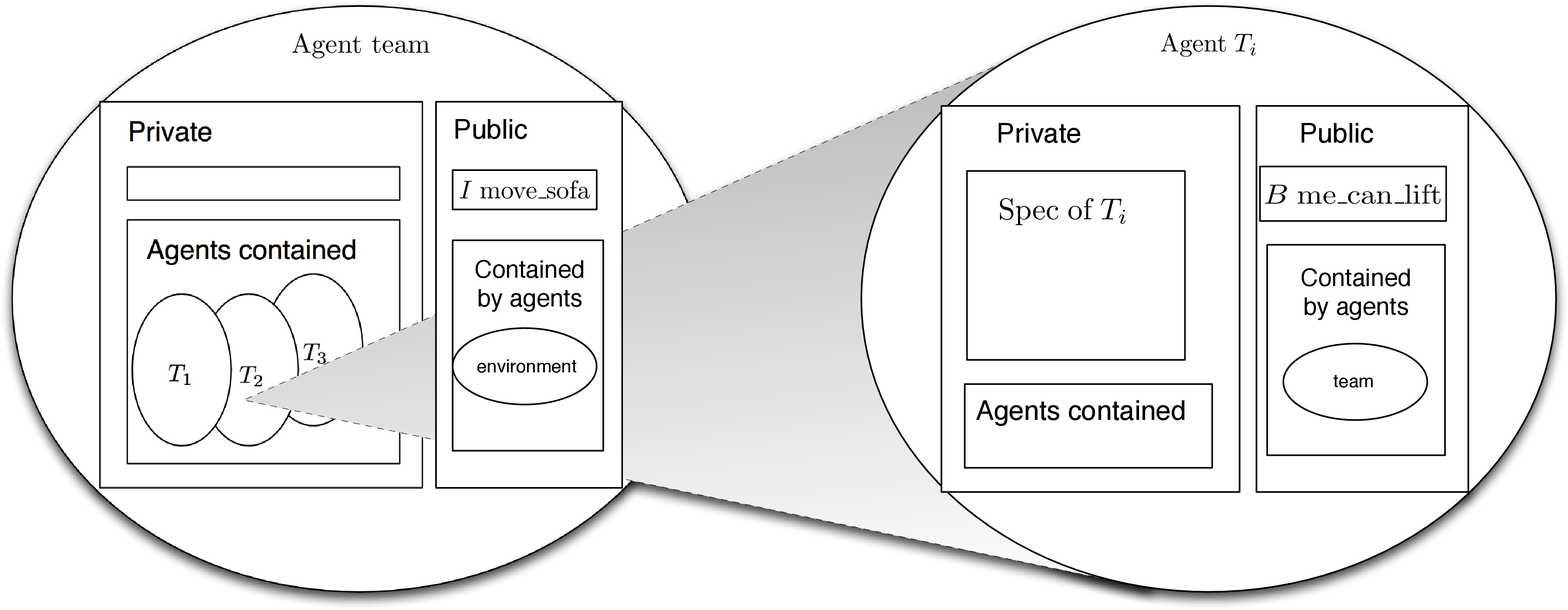}
                \caption{Representing a team using the agent
                  structure~\cite{FisherDH:2009}.}
                        %\vspace{-0.4cm}
               \label{fig:exteam}
               \vspace{-0.6cm}
        \end{figure}

In~\cite{FisherDH:2009,HeppleDF:2008} this agent specification
approach is extended further to include an agent's specification that
is visible, or accessible, to the agent's content or context
respectively. Effectively this means that one part of the agent is
visible to the agent's content and another part, possibly the same, is
visible to the agent's context. This extension allows for the
``typical'' agent structures, in which the agents have joint or shared
attitudes, to be represented. We give a simple example of a team using
this representation in Fig.~\ref{fig:exteam}. The agent
$\mathit{team}$ has, in its content, all the agents $T_1, T_2,
T_3$. The team agent has we-attitudes visible to its content, in
this example it is the intention to ``move the sofa'', but nothing in
its private specification. The agents $T_i$ have their own shared
attitudes, in this example the belief that the agent in question can
``lift'' (the sofa), visible to the context and in the private
specification, the rest of their mental attitudes.

The \emph{context}/\emph{content} extension enables an agent structure
to be directly represented but, regardless of whether the agent
specification is private or public, it is still in a typical BDI
specification. What all BDI-oriented agent specification languages
have in common is that they express what the agent ``thinks'' and how
it ``reasons''.  Although the agent might choose to reveal some of its
mental attitudes, this is still internal information. The information
that the agent wants to share with the environment and other agents
might not correspond to, or even be consistent with, its formal
specification.
%% In a crowd, an agent might have its public information made
%% available, but this does not mean that we do not know anything
%% about that agent, we still know that he decided to join a certain
%% platform, or a crowd and we know what information it can access,
%% which tells us about the incentives to which the agent may respond.  
To model a digital crowd participant we also need to model what the
agent ``hears'' and ``says'', potentially hiding what the agent
``thinks'' from other agents.
%Namely, we need to specify the information that the agent chooses to share with its environment.  Furthermore, one has to bear in mind that crowds are characterised with an immensely larger number of participants, compared with traditional agent organisations, and detailed representations of the mental state of each and every crowd participant is not only unnecessary, but also computationally unfeasible. 
Thus, to distinguish what the agent ``was told'' and ``has said'' from
its own private ``thought'' behaviour we extend the above approach and
explicitly represent messages that are sent to (and received from) the
agent's content or context. By having this information explicitly
represented, the agent can reason about it and react to it
accordingly. 
%% The different ways in which the agent reasons about, or react to,
%% received or sent information would specify different behaviors.
 
\subsection{Communication as an informational state}
A formal specification of an agent is a logic formula that describes
its behaviour, including the information it has about the world, the
motivations that drive it, and the methods that define how it reacts
in response to messages. Logics developed for representing different
aspects of agent behaviour include modal logics of belief ($B$), goals
($G$), wishes ($W$), desires ($D$), intentions ($I$), actions
($\lsem\mathcal{A} \rsem$), abilities ($A$), knowledge ($K$), etc, all
with an underlying temporal and/or probabilistic
basis~\cite{schild,Practical11}. Following this tradition, we
introduce a new modal logic operator ``$M$'' to handle message
passing, representing messages as $M \varphi$, where $\varphi$
describes the message content. We here consider that the message content is a well formed formula for the language used to represent the mental states of the agent. 
 
There are two basic qualifiers that must be attributed to a message to
distinguish between messages that have been received and messages that
have been sent. Additionally, it is also necessary to identify the
sender or recipient, respectively.  To accomplish this distinction we
modify the $M$ operator using a superscript $\uparrow\!  j$ denoting
messages sent to agent $j$ and $\downarrow\! j$ denoting messages
received from agent $j$. Thus, given an agent communication
specification, including $M^{\uparrow j} \varphi$ in the specification
requires that the agent has sent a message with content $\varphi$ to
agent $j$. Similarly, $M^{\downarrow j} \varphi$ represents that the
specified agent has received a message with content $\varphi$ from
agent $j$. However, the modifiers $\uparrow\!  j$ and $\downarrow\! j$
are insufficient to express complex messages such as incentives.

We further modify the $M$ operator with a {\em type}, denoted as a
subscript \eg $M^{\uparrow j}_{type} \varphi$, clarifying the intended
meaning of the message. We distinguish between four types.
\begin{itemize}
\item[]\begin{itemize}
\item[{\bfsl tell}] denotes information passing, implying the sender
  informs the receiver of what its mental state is; for example,
  $M^{\uparrow j}_{tell} I \varphi$ is a message with which the
  specified agent informs agent $j$ of its intention to actively
  pursue $\varphi$ (i.e., it has $\varphi$ as an intention).
\item[{\bfsl ask}] denotes a request for confirmation of state; for
  example, $M^{\downarrow j}_{ask} I \varphi$ represents the fact that
  the agent has received an inquiry from $j$ who wants to know whether
  the agent intends to pursue $\varphi$.
\item[{\bfsl do}] denotes delegation, extension and transfer of
  attitudes; for example, $M^{\downarrow j}_{do} I \varphi$ is a
  message in which $j$ delegates to the recipient the active pursuit
  of $\varphi$, while $M^{\uparrow j}_{do} B \varphi$ is a message in
  which the sender extends the requirement of adopting the belief
  $\varphi$ to agent $j$, etc. (Note that, to send or receive a
  message of delegation does not mean that the recipient is obliged to
  adopt the content.)
\item[{\bfsl adv}] denotes announcement, promise and advertisement, and
  allows agents to inform others of incentives or constraints and
  other mental attitudes that are in force for prospective members;
  for example, $M^{\uparrow j}_{adv} I \varphi$ denotes that if the
  recipient $j$ joins the content of the sender, then $i$ will be
  asked to adopt the intention $\varphi$, while $M^{\downarrow
    j}_{adv} I \varphi$ denotes that if $j$ is added to the content of
  the recipient, then $j$ will adopt the intention $\varphi$.
\end{itemize}\end{itemize}
Table~\ref{tab:intmes} gives the intuitive interpretation
of such communication types in the specification of agent $i$.

\begin{table}[h!]
\centering
\begin{tabular}{|c|c|} \hline
\multicolumn{2}{|c|}{\bfsl tell} \\\hline
{\large $M^{\downarrow j}_{tell} \varphi$}& {\large $M^{\uparrow j}_{tell}\varphi$} \\ 
&\\
$i$ is told that $\varphi$ holds for $j$ & $j$ is told that $\varphi$
holds for $i$\\\hline \multicolumn{2}{c}{\ }\\ \hline

\multicolumn{2}{|c|}{\bfsl ask}\\ \hline
{\large $M^{\downarrow j}_{ask}\varphi$}& {\large $M^{\uparrow j}_{ask}\varphi$}\\ 
& \\ 
$j$ asks: does $\varphi$ holds for $i$?& $i$ asks: does $\varphi$ holds for
$j$?\\\hline\multicolumn{2}{c}{\ }\\ \hline

 \multicolumn{2}{|c|}{\bfsl do}\\\hline
{\large $M^{\downarrow j}_{do}\varphi$}& {\large $M^{\uparrow j}_{do}\varphi$} \\ 
& \\ 
$j$ says $\varphi$ should hold for $i$ & $i$ says $\varphi$ should
 hold for $j$ \\\hline\multicolumn{2}{c}{\ }\\ \hline

\multicolumn{2}{|c|}{\bfsl adv}\\ \hline
{\large $M^{\downarrow j}_{adv}\varphi$}& {\large $M^{\uparrow j}_{adv}\varphi$} \\
& \\ 
$\varphi$ should hold for $i$,& $\varphi$ should hold for $j$, \\
 if $i$ were in the content of $j$ & if $j$ were in the content of $i$
\\ \hline

\end{tabular}\caption{Intuitive interpretation of the messages in the  specification of agent $i$. Agent $j$ is an agent different from agent $i$.}\label{tab:intmes}
\end{table}
\medskip
\noindent\textbf{Notation.} In the rest of the text, we use the
notation $M^{\uparrow}_{[\cdot]} $ and $M^{\downarrow}_{[\cdot]} $,
respectively to denote messages of any type, and we use the
$\downuparrows$ symbol when the orientation of the message (incoming
or outgoing) is irrelevant.
\medskip

\noindent The above message type modifiers are inspired by the
illocutionary (speech) act types of~\cite{Searle1975}. Searle
distinguishes among assertives, directives, commissives, expressives
and declarations. We can draw correspondences between our message
types and Searle's illocutionary types: expressives and messages of
type {\sl tell}, assertives and messages of type {\sl ask}, directives
and messages of type {\sl do}, and commissives and messages of type
{\sl adv}. We do not have a message type that corresponds to
declarations, which are speech acts that permanently change the world,
such as pronouncing someone as being married, employed etc, because we
do not consider these types necessary for representing digital crowds
at present.

Within MAS research, agent languages for \mfchange{the} coordination
of agents, based on speech acts, have been developed, \eg
~\cite{VieiraMWB:2007,BarbuceanuF:1995}. However, communicating using
these languages, the agents use speech acts, with vocabulary and
interpretation of the received messages depending on the specific
language used. As \mfchange{discussed} in Section~\ref{problem}, our
messages are not speech acts, rather, they are the agent's attitudes
and the illocutionary act types are used to modify the intended
meaning of the communication, the reason why the message content is
sent, or how the message content is to be interpreted.
 
Logics have been developed that are concerned with exchange of
knowledge among agents, probably the most notable being {\em
  dynamic epistemic logic}~\cite{Ditmarsch:2007}. It is important to
outline the difference between our $M$ operator and the communication
represented in this logic.  Within dynamic epistemic logic, one can
specify information that is communicated to all the agents, called an
{\em epistemic update}, as well as knowledge that all the agents
possess, \ie common knowledge. Once an update $\varphi$ occurs, the
agent that receives it necessarily changes her mental states to
accommodate $\varphi$. As a result, from the moment of the update
onwards the agent believes that $\varphi$ holds. However, when an
agent receives a message $M \varphi$, she has a choice of whether to
accommodate $\varphi$ in her mental states or not. In other words,
after a message $M^{\downarrow i}_{tell} B \varphi$ or even after a
message $M^{\downarrow i}_{do} B \varphi$ is received, it is not
necessary that the agent will have $B \varphi$ as part of her
specification. Explicitly, we do not presume that the agent tells the truth in the content of  a ``tell" message. Thus,  $M^{\uparrow i}_{tell} B \varphi$ and $B \neg \varphi$ are mutually consistent. 
 
\section{Logically Specifying Agents}\label{s:specification}
Let $Agt $ be a set of unique agent identifiers, let $Prop$ be a set
of atomic propositions \mschange{and constants}, and $Pred$ be a set of a first-order predicates
of arbitrary arity. We begin by defining a language $\mathcal{L}_p$ to
be a set of grounded first order logic formulas without function
symbols, namely the set of all $\varphi_p$ such that
\[  \varphi_p::=   p \mid \neg \varphi_p \mid \varphi_p \wedge \varphi_p \mid P(x_1, \ldots, x_m) \]
 where $p\in Prop$, $P \in Pred$  and $x_1, \ldots, x_m \in Agt$.

We next consider a $\mathit{BDI}$ agent language. Depending on the
specific needs for a specification, different $\mathit{BDI}$ operators
can be used but, for demonstrating our specification
\mfchange{approach}, we use the modal operators $B$, $G$ and $I$, to
denote agent's beliefs, long term goals and actively pursued goals 
\mfchange{(or intentions)}, respectively. We also use an operator,
$A$, to denote that an agent has an ability, $A \varphi$ indicating
that the agent is able to accomplish $\varphi$. Ability is a more
complex mental attitude and several formalizations are possible, \eg
``ways''~\cite{Elgesem:1997,BelnaoP:1988,Troquard:2013,WobckePZ:1998},
depending on the precise interpretation of ability.  To avoid an
in-depth philosophical analysis of the logics of abilities, which is
outside of the scope of this work, and to simplify our modelling
language by avoiding the use of actions as logical primitives, we
resolve to use $A \varphi$ for denoting both procedural knowledge of
which action sequence brings about $\varphi$ and the actual ability to
perform an action sequence that brings about $\varphi$. The language
$\mathcal{L}_{\mathit{BDI}}$ is then the set of all formulas $\varphi$
such that
\[\varphi :: =    \varphi_p \mid  \neg \varphi  \mid \varphi \wedge \varphi  \mid B  \varphi_p \mid I \varphi_p \mid G\varphi_p \mid A\varphi_p, \]
where $\varphi_p \in \mathcal{L}_p$.

\mfchange{Finally}, we define the language for specifying
communication among agents, $\mathcal{L}_M$. For this language,
temporal logic operators should be specified depending on the needs of
the particular system specified. We use $\mathit{LTL}$ operators in
our examples~\cite{Practical11}. Let $T$ be the set of message types
$T=\{tell, ask, do, adv\}$ and $Agt$ be a set of unique agent
identifiers. The language $\mathcal{L}_M$ is the set of all formulas
$\theta$ such that
\[  \theta::=   \varphi  \mid   \neg \theta \mid  \theta  \wedge \theta 
  \mid \varphi {\bf U } \varphi \mid \bigcirc \varphi \mid \Diamond
  \varphi \mid M^{\downarrow j }_{x} \theta \mid M^{\uparrow j}_{x}
  \theta, \] 
where $\varphi \in \mathcal{L}_{\mathit{BDI}}$ and $x \in T$. In
the intuitive interpretation of temporal operators: $p{\bf U }q$ means
that $p$ is continuously true up until the point when $q$ becomes
true; $\bigcirc r$ means that $r$ is true in the next moment in time;
while $\Diamond s$ means that $s$ will be true at some moment in the
future.

The messages are sent to an agent $j$, however either the
\emph{context} set $CX$ or the \emph{content} set $CN$ as a whole can
be the target of message broadcast. We use the
shorthand\footnote{\textbf{Note:} We define the messages with
  individual agents, not sets as in
  \cite{Fisher99,FisherDH:2009,HeppleDF:2008}, because a message can
  be broadcast to many agents, but it can be sent from one agent,
  otherwise the sender is unknown, which \mfchange{cannot happen here}
  --- if your contexts sends you a message it is from exactly one
  context.}

%\[  M^{\downarrow CN} _{type} \varphi \equiv \bigwedge_{i\in CN}   M^{\downarrow i}_{type} \varphi,\;\;\;  M^{\downarrow CX} _{type} \varphi \equiv \bigwedge_{i\in CX}   M^{\downarrow i}_{type} \varphi \]
\[  M^{\uparrow CN} _{[\cdot]} \varphi \equiv \bigwedge_{j\in CN}   M^{\uparrow j}_{[\cdot]} \varphi,\;\;\qquad\;  M^{\uparrow CX} _{[\cdot]} \varphi \equiv \bigwedge_{j\in CX}   M^{\uparrow j}_{[\cdot]} \varphi. \]

\noindent The language $\mathcal{L}_{\mathit{BDI}}$ restricts the
nesting of modal operators, while $\mathcal{L}_{M}$ forbids the use of
$\mathit{BDI}$ and temporal operators outside of the scope of a
message operator. Thus the agents do not have mental attitudes about
the future, \eg $B \sometime \varphi$ ``I believe that sometimes
$\varphi$ is true" nor $B G \varphi$ ``I believe I have the goal to
accomplish $\varphi$".

Nested messages express meta communication, allowing agents to
communicate about what was communicated to them or by them. However
not all nesting is meaningful.  We state constraints as shown in
(\ref{nesting}) in order to apply restrictions mandating that all but the innermost and outermost
message operators are dropped, the orientation of the inner most and
outermost messages is retained, as is the type of the innermost
operator, while the type of the outermost operator must be $tell$.
   \begin{equation}\label{nesting}
   \begin{array}{c}
   M^{\downarrow i}_{[\cdot]} M^{\downuparrows i}_{[\cdot]}  \cdots     M^{\downuparrows i}_{[\cdot]}  M^{\downarrow i}_{[x]} \varphi \leftrightarrow    M^{\downarrow i}_{[tell]} M^{\downarrow i}_{[x]} \varphi \\ 
   \;\\
      M^{\downarrow i}_{[\cdot]} M^{\downuparrows i}_{[\cdot]}  \cdots     M^{\downuparrows i}_{[\cdot]}   M^{\uparrow i}_{[x]} \varphi \leftrightarrow    M^{\uparrow i}_{[tell]}  M^{\uparrow i}_{[x]} \varphi  \\ 
         \;\\
   M^{\uparrow i}_{[\cdot]} M^{\downuparrows i}_{[\cdot]}  \cdots     M^{\downuparrows i}_{[\cdot]}  M^{\downarrow i}_{[x]} \varphi \leftrightarrow    M^{\uparrow i}_{[tell]}M^{\downarrow i}_{[x]} \varphi  \\  
      \;\\
      M^{\uparrow i}_{[\cdot]} M^{\downuparrows i}_{[\cdot]}  \cdots     M^{\downuparrows i}_{[\cdot]}  M^{\uparrow i}_{[x]} \varphi \leftrightarrow    M^{\uparrow i}_{[tell]}  M^{\uparrow i}_{[x]}\\       
   \end{array}
   \end{equation}
\noindent We assume that (some of) the agent abilities are
transferable and thus messages such as $M^{\downarrow i}_{do} A
\varphi$ and $M^{\uparrow i}_{do} A \varphi$ represent the transfer of
abilities from agent $i$ to the sender and vice versa,
respectively. Messages such as $M^{\uparrow i}_{adv} A \varphi$
represent transfer of abilities after the agent $i$ has joined the
content of the sender.
\smallskip

\noindent We can now give the following definition of an agent. 
\begin{definition}
Let $Agt$ be a set of unique agent identifiers.  An agent is a tuple
\linebreak $\langle \mathit{ID}, Bel, Int, Goal, Ablt, Com, CN, CX
\rangle$, where $\mathit{ID} \in Agt$ is a unique agent identifier,
$Bel \subset \mathcal{L}_p$ is the set of beliefs the agent holds
about the world, $Int \subset \mathcal{L}_p$ is the set of
\mfchange{the agent's} intentions, $Goal \subset \mathcal{L}_p$ is the
set of the agent's goals, $Ablt\subset \mathcal{L}_p$ is the set of
the agent's abilities, $Com \subset \mathcal{L}_M$ is the set of
messages the agent has received and sent, $CN \subset \mathcal{P}( Agt
\setminus \{\mathit{ID}\})$ is the set of agents contained and lastly
$CX \subset \mathcal{P}( Agt \setminus \{\mathit{ID}\})$ is the set of
agents in which the agent is contained, \ie its set of contexts. Each
of the sets $Bel$, $Int$, $Goal$ and $Ablt$ are consistent and
simplified.

Given an agent $i \in Agt$, an agent specification is a set
$\mathit{SPEC}(i) \subset \mathcal{L}_M$, where $B \varphi$ is true
iff $\varphi \in Bel$, $G\varphi$ is true iff $\varphi \in Goal$, $I
\varphi$ is true iff $\varphi \in Int$, $A \varphi$ is true iff
$\varphi \in Ablt$, $cn(j)$ is true iff $j \in CN$, $cx(j)$ is true
iff $j\in CX$ and $M^{\downuparrows i}_{[\cdot]} \varphi$ is true if
$M^{\downuparrows i}_{[\cdot]} \varphi \in Com$.
\end{definition}

\noindent Note that, to be able to reason about contents and contexts,
we introduce special formulas in $\mathit{SPEC}(i)$, namely $cn(j)$
and $cx(j)$. The formula $G\; cx(j)$ within $\mathit{SPEC}(i)$
expresses $i$'s goal to have $j$ as a context agent, and similarly
$G\; cn(j)$ within $\mathit{SPEC}(i)$ denotes the goal to include $j$
in his content. Symmetrically $G\; \neg cx(j)$ and $G\; \neg cn(j)$
express the goal to remove $j$ from ones own context, or content,
respectively.
 
% We can introduce special propositions $jx$, $jn$,  and $leave$ to express joining or leaving a context/content. Additionally we can introduce special predicates:
%\begin{itemize}
%\item $begin\_\alpha$ - I begun doing some action$\alpha$; 
%\item $end\_\alpha$ - I finished doing some action $\alpha$;
%\item $doing\_\alpha$- I am currently executing action $\alpha$.
%\end{itemize}

%  We also introduce special (grounded) predicates  $cn(i)$ : true iff $i \in CN$, 
% $cx(j)$ : true iff $j\in CX$, and  $agt(i)$: true iff $i \in Agt$. 

Lastly, we assume, via (\ref{connectivity1}), that if a message is
sent then it will eventually be received. This is a property of
communication among agents that should hold in the environment, for
communication to be meaningful.
 \begin{equation}\label{connectivity1}\exists i, M^{\uparrow j}_{[\cdot]} \varphi \in \mathit{SPEC}(i)   \Rightarrow  \exists j , \sometime M^{\downarrow i}_{[\cdot]} \varphi \in \mathit{SPEC}(j) \end{equation}
% 
% \item If a message is received from $j$, then $j$ sent it at some point  in the past. 
% \begin{equation}\label{connectivity2}\exists i, M^{\downarrow j}_{[\cdot]} \varphi \in \mathit{SPEC}(i) \Rightarrow  \exists j , \somepasttime M^{\uparrow i}_{[\cdot]} \varphi \in \mathit{SPEC}(j) \end{equation}
 
\noindent Note that we do not develop an axiomatisation for
$\mathcal{L}_M$ and do not intend to prove soundness for this
language, because we intend ultimately to use it to create
specifications for model checking, where soundness is not necessary.
The above, together with standard modal and temporal logic semantic
structures~\cite{Stirling89}, provides a formal basis for describing
digital crowd structures, communication and, hence, behaviour.

\section{Specification and Verification of Digital Crowds}
\label{examples}
We demonstrate the applicability of the specification notation above
by considering two digital crowd examples, and describing the
verification processes that can be deployed.

Throughout, we assume that verification tools for individual agents,
such as those available to analyse the $\mathit{BDI}$ properties of
\texttt{Java}-based agents~\cite{MCAPL_journal}, can be utilised to
assess individual agents. We here show, using examples, how the
specification language we have developed is appropriate for describing
and reasoning about the crowd behaviour, and that properties we might
wish to establish of the crowd can be built up from the properties of
individual agents. Note also that, while the proofs we outline are
provided by us, we would expect automated (or, at least,
semi-automated) provers to be able to generate these in a
straight-forward way.
\smallskip

\noindent We begin by discussing some common aspects of digital crowds.
\smallskip

%For building our specification, we  use the modal operators $B$, $G$ and $I$, to denote agent's beliefs, long term goals and actively pursued goals, or intentions, respectively. We  also use an operator $A$ to denote that an agent has an ability, $A \varphi$ denoting that the agent is able to accomplish $\varphi$. Ability is more complex mental attitude and several  formalizations are possible, \eg   ways  \citep{Elgesem:1997,BelnaoP:1988,Troquard:2013,WobckePZ:1998}, depending on the precise interpretation of ability.    
%
%To avoid an in-depth philosophical analysis of the logics of abilities, which is outside of the scope of this work, and simplify our modeling language, by avoiding to use actions as logical primitives, we resolve to use $A \varphi$ for denoting both procedural knowledge of which action sequence brings about $\varphi$ and  actual ability to perform an action sequence that brings about $\varphi$. Further, we assume that abilities are transferable and thus messages  such as $M^{\downarrow i}_{do} A \varphi$ and  $M^{\uparrow i}_{do} A \varphi$ represent the transfer of abilities from agent $i$ to sender and vice versa, respectively. Messages such as   $M^{\uparrow i}_{adv} A \varphi$ represent transfer of abilities after the agent $i$ has joined the content of the sender. 
\noindent In a (digital) crowd, the agents are highly autonomous with
goals, and attitudes in general, beyond a seeker's control. As a
consequence, agents cannot be given a goal, they can only be inspired,
or be provided with incentives, to adopt a goal themselves. In
economics, an {\em incentive} is considered to be a cost or benefit
that motivates decisions or actions of agents. As observed in the
multitude of work concerned with crowdsourcing,
\eg~\cite{Estelles-Arolas:2012,HoZWVS:2012,KamarH:2012,SchenkG:2011},
one of the key features of the \cws process is the provision of
incentives, by the seeker, in the form of benefits to the agents that
(successfully) participate in the \cws process.  We can specify, as an
advertisement, a conditional future possibility to benefit with the
schema (\ref{def:incentive}).
\begin{equation}\label{def:incentive}
[INC]:\quad  M^{\updownarrows i}_{adv} (\mathit{ conditions } \rightarrow \sometime \mathit{ reward } ).
\end{equation}

\noindent The incentive (\ref{def:incentive}) is a message of type
``promise''.  The premise $\mathit{conditions}$ of the message is a
formula describing the required skills, the pursuit of required goals,
etc., that could lead to benefit in the content of the sender.  The
$\mathit{reward}$ is the benefit that an agent can obtain, such as
money, goods, recognition, status, etc. In our examples we use the
formula ``$A \; \mathit{earn}$'' as a general representation for an
earned reward.

\mfchange{In} the same manner as specifying positive incentives, those
that promise gain, we can \mfchange{also} specify negative incentives
that promise a penalty. In this case, the conditions specify what the
agent should refrain from, \eg intentions not to be upheld, and
instead of $\sometime \mathit{ reward} $ as a consequent we would have
$\sometime \mathit{ penalty} $. While penalties are of interest in
agent structures such as institutions, in crowdsourcing, penalties are
not commonly utilized; therefore we investigate them no further here.

We cannot, nor do we want to, access the mental states of each of the
agents of the crowd, but we can look at the exchanged messages and,
based on assumptions about how an agent interprets those messages, we can
establish certain properties of a crowd.

\subsection{Information Retrieval}
We next consider an example in which the retrieval of \mfchange{a}
particular item of information is \emph{crowdsourced}, such as finding
a suspect's name (in \mfchange{our} case, $\mathit{Nemo}$) from an
image in a criminal investigation. We want to formally ascertain that
Nemo will be found under the assumption that there exists an agent in
the crowd who knows where Nemo is, as long as some simple assumptions
for the relations among the agents in the crowd hold as well.

Note that, making assumptions about a system and then verifying that
under these assumptions certain properties hold is not unusual in
\mfchange{formal} verification. While the exact state of a system
cannot always be known, assuming that the system is in \mfchange{the
  required} state, we can \mfchange{ensure} that a desired state can
be reached.
%% \texttt{In the sentence above I would like to explain that the goal
%% of verification is not to determine the exact behavior of the
%% system, but to prove that set boundaries on the behavior will be
%% maintained.  I am certain this can be done better.}
This allows us to be certain that, when operational conditions allow,
correct behaviour will ensue; it also allows us to explore assumptions
concerning failure in any of these operational conditions --- if
\mfchange{the pre-requisites} we expect \mfchange{are} not actually
present, then what can the behaviour of the system be?

In order to consider communication among crowds, let us define the
concept of {\em reachability} between two agents $i$ and $j$.  The
agent $i$ can reach agent $j$ if, and only if, a message sent from $i$
is eventually forwarded to $j$, under the assumption that the relevant
context relay messages from one of their content agents to the rest of
the content.  We first define \emph{relaying contexts}. Intuitively, a
relying context is an agent which broadcasts to all its content agents
all messages received from one of his content agents.

\begin{definition}
Let $i$ be an agent s.t. $CN(i) \neq \emptyset$. Agent $k \in CX(i)$
is a {\em relaying context}, \mfchange{and} $\mathit{REL}(k)$ is true,
when all the messages sent to $k$ are sent \mfchange{on} to all of the
content agents of $k$:
\[ ((CN(i)\vee CX(i)) \wedge M^{\downarrow i} _ {tell} \varphi) \rightarrow M^{\uparrow CN} _ {tell} \varphi )\in \mathit{SPEC}(k) \]
\end{definition}

\noindent Clearly, there are messages that an agent sends to a content
that are not to be shared, and should be kept private, but these can
be specifically designated as such.  The reachability between two agents is now defined
recursively.
\begin{definition}
Agent $j$ is {\em directly reachable} for agent $i$ if at least one of the following is true:
\begin{itemize}
\item $\exists k\in CX(i) \cap CX (j) $ s.t.  $\mathit{REL}(k)$ holds.
\item  $ \exists k_1 \in CX(i) $ and   $ \exists k_2 \in CX(j) $ s.t. $CN(k_1) \cap CN(k_2) \neq \emptyset$ and $\mathit{REL}(k_1) \wedge  \mathit{REL}(k_2)$ holds.
\end{itemize}
Agent $j$ is {\em  reachable} for agent $i$ if at least one of the following is true:
\begin{itemize}
\item $j$ is {\em directly reachable} for agent $i$ .
\item  $\exists k \in Agt$  s.t. $k$ is reachable for $i$ and $j$ is reachable for $k$.
\end{itemize}

\end{definition}
 
\begin{figure}
  \vspace{-0.5cm}
        \centering
       \includegraphics[width=0.99\textwidth]{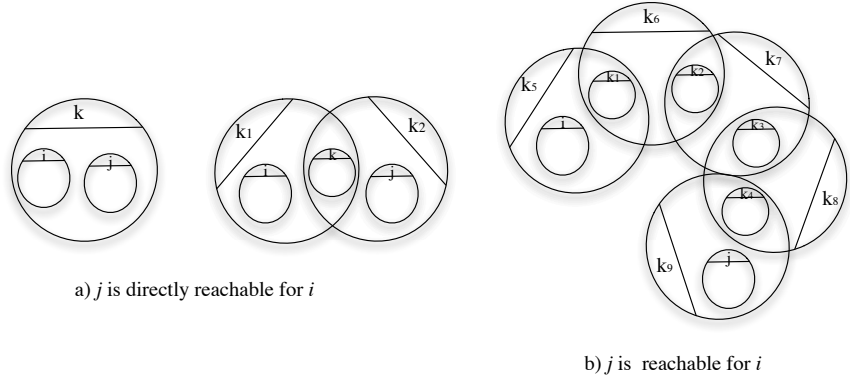}
                \caption{Examples of direct reachability and reachability.}
                \label{fig:agent}
                \vspace{-0.5cm}
        \end{figure}

\noindent In Fig.~\ref{fig:agent}, we give an example of directly
reachable (Fig.~\ref{fig:agent}(a)) and reachable
(Fig.~\ref{fig:agent}(b)) agents.
 
The set of all agents that are reachable for $i$ is called the
neighbourhood of $i$, $\mathit{NGH}(i)$.  Determining whether $j \in
\mathit{NGH}(i)$, under the assumption that all context agents are
relaying contexts, can be solved as \mfchange{an} ST-connectivity
problem~\cite{papadimitriou94}.

There are always certain assumptions that have to be made concerning
the behaviour of members of a digital crowd. We will make the
assumptions that the following formulas are included in the
$\mathit{SPEC}$ of every agent in $Agt$:

\begin{enumerate}
\item[$\ast$] An agent is always pursuing the goal to earn (increase
  \mfchange{her} utility):
  \begin{equation}\label{asumption1}
     G \; \mathit{ earn }.
  \end{equation}

\item[$\ast$] An agent always shares information with its content and
  context about any possibilities to earn, namely incentives are
  forwarded:
  \begin{equation} \label{asumption2}M^{\downarrow j}_{adv}  \mathit{INC} \rightarrow M^{\uparrow CN}_{tell} M^{\downarrow j}_{adv}  \mathit{INC}. 
  \end{equation}
  \begin{equation}\label{asumption3} M^{\downarrow j}_{adv}  \mathit{INC} \rightarrow M^{\uparrow CX}_{tell} M^{\downarrow j}_{adv}  \mathit{INC}.
  \end{equation}

\item[$\ast$] A forwarded incentive is treated the same as directly
  received incentive:
  \begin{equation} \label{asumption4} M_{tell}^{\downarrow k} M^{\downarrow j}_{adv}  \mathit{INC}\rightarrow M^{\downarrow j}_{adv}  \mathit{INC}.
  \end{equation}

\item[$\ast$] An agent always responds to an incentive whose
  conditions it can satisfy:
  \begin{equation} \label{asumption5} (\mathit{conditions }\wedge G\mathit{ reward } \wedge  M^{\downarrow j}_{adv} (\mathit{conditions} \rightarrow \sometime  \mathit{ reward } )) \rightarrow M^{\uparrow j}_{tell} \mathit{ conditions}.
  \end{equation}

\item[$\ast$] An agent is required to have the abilities it claims to
  have, hence it will tell someone that it has a certain ability 
  only if it really has it:
  \begin{equation} \label{asumption6} 
   M^{\uparrow j}_{tell} A \varphi \rightarrow A \varphi.
  \end{equation}

\item[$\ast$] An agent that actively pursues a goal, and has the
  ability to accomplish its goal, will eventually (believe that
  \mfchange{she} will) have accomplished his goal:
  \begin{equation} \label{asumption7}  
     (A \varphi \wedge I \varphi)\rightarrow \sometime B \varphi.
  \end{equation}

\end{enumerate}
Assumptions about the behaviour of the crowd can be based on a
statistical analysis of the agents that access a platform of interest,
or by only allowing agents with desirable properties to access the
platform.  Thus, a condition that all contexts are relaying messages
can be established by ensuring a platform in which all communication
among the members of a structure is visible to all others (consider as
an example a \emph{Facebook wall}). \mfchange{It is important to note that, if we move to a different infrastructure, we might modify or even remove some of these assumptions. The key point is that the specification formalism is appropriate for describing these.}

In addition to the general assumptions
(\ref{asumption1})-(\ref{asumption7}) we make the following
\mfchange{specific} assumptions for $\mathit{SPEC}(seeker)$ within
this particular information retrieval scenario.

\begin{enumerate}
\item[$\ast$] If an incentive was advertised that required
  \mfchange{the} ability $\mathit{find\_Nemo}$ and an agent tells us
  that it has such an ability, then delegate the task of
  $\mathit{find\_Nemo}$ to that agent:
  \begin{equation}\label{assumption8}
(M^{\uparrow CX}_{adv} (A\; \mathit{find\_Nemo} \rightarrow \sometime A\; \mathit{ earn }) \wedge M^{\downarrow k}_{tell} A\; \mathit{find\_Nemo}) \rightarrow M^{\uparrow k}_{do} I\; \mathit{find\_Nemo}.
  \end{equation}

\item[$\ast$] If a task to   $\mathit{find\_Nemo}$ was delegated to  an agent $k$, and a message was received from $k$ that Nemo is found, then  the belief that Nemo is found is adopted:
  \begin{equation}\label{assumption8a}
(M^{\uparrow k}_{do} I\; \mathit{find\_Nemo} \wedge M^{\downarrow k}_{tell}  B\; \mathit{find\_Nemo}) \rightarrow B\; \mathit{find\_Nemo}. 
  \end{equation}
\end{enumerate}
Note that, because of (\ref{assumption8}), all agents that have reported
an ability to $\mathit{find\_Nemo}$ will be assigned the task to
$\mathit{find\_Nemo}$. From (\ref{assumption8a}), it follows that as soon as one of
the delegated agents reports that Nemo is found, the respective belief
will be adopted.
 \medskip

\noindent We also make additional assumptions for the $\mathit{SPEC}$
of the agents in $Agt.$
\begin{enumerate}
\item[$\ast$] If we had told an agent that we have the ability to
  $\mathit{find\_Nemo}$, and that agent delegated to us the finding of
  Nemo, then we will indeed adopt the intention to
  $\mathit{find\_Nemo}$:
  \begin{equation}\label{assumption9}
(M^{\uparrow i}_{tell}A\; \mathit{find\_Nemo} \wedge   M^{\downarrow i}_{do} I\; \mathit{find\_Nemo} ) \rightarrow   I\; \mathit{find\_Nemo}. 
  \end{equation}

\item[$\ast$] If we believe that \mfchange{the} $\mathit{find\_Nemo}$
  \mfchange{task has been} accomplished, and we were delegated by an
  agent to find Nemo, then we tell that agent that we believe Nemo is
  found:
  \begin{equation}\label{assumption10}
(B\; \mathit{find\_Nemo} \wedge  M^{\downarrow i}_{do} I\; \mathit{find\_Nemo} ) \rightarrow M^{\uparrow i}_{tell}  B\; \mathit{find\_Nemo}.  
  \end{equation}
\end{enumerate}
We aim to verify whether property (\ref{seeker}) holds for agent
$seeker$ when the platform is such that property (\ref{crwagt}) is
satisfied ($Agt$ is the set of all agents in the platform). Property
(\ref{crwagt}) expresses that there is at least one agent in the
neighbourhood of the Seeker that is able to find Nemo. Recall that the
communication among agents is something that can be accessed by other
agents, therefore we establish that an agent has an ability if it has
said it has.
\begin{enumerate}
\item[$\ast$] There is at least one agent $j$ in the neighbourhood of
  the $seeker$ that has communicated possessing the ability to
  $\mathit{\mathit{find\_Nemo}}$ to some agent $k$, or was told by $k$
  that $k$ has the ability to $\mathit{\mathit{find\_Nemo}}$:
  \begin{equation}\label{crwagt}
\exists j\in \mathit{NGH}(seeker),\ M^{\uparrow k}_{tell} A \;\mathit{\mathit{find\_Nemo}} \in \mathit{SPEC}(j) \mbox{ or }  M^{\downarrow k}_{tell} A \;\mathit{\mathit{find\_Nemo}}\in \mathit{SPEC}(j).
  \end{equation} 

\item[$\ast$] It is always true that if we advertise the incentive
  that the ability to find Nemo can lead to a reward (arbitrarily long), then eventually
  Nemo will be found:
\begin{equation}\label{seeker}
 \always (M^{\uparrow seeker}_{adv} (A \;\mathit{\mathit{find\_Nemo}} \rightarrow \sometime A\; { earn })  \rightarrow  \sometime B\; \mathit{find\_Nemo}).
\end{equation}
\end{enumerate}
Let $j$ be the agent that is in $\mathit{NGH}(seeker) $
s.t. $M^{\uparrow k}_{tell} A \;\mathit{\mathit{find\_Nemo}} \in
\mathit{SPEC}(j)$, namely the agent that makes (\ref{crwagt})
true. Following from (\ref{asumption4}) and (\ref{asumption5}), we have that
$j$ (in $\mathit{SPEC}(j))$ it will eventually obtain the advert from
the $seeker$:
 \[\sometime M^{\downarrow \mathit{seeker}}_{adv} (A \;\mathit{\mathit{find\_Nemo}} \rightarrow \sometime A\; { earn })  .\]
\mfchange{Since  (\ref{crwagt}) \mschange{is} true, and due to (\ref{asumption6}),}
 (\ref{asumption1}) and (\ref{asumption5}), the $seeker$ will
 eventually be contacted by $j$, namely $\mathit{SPEC}(seeker)$
 contains
  \[\sometime M^{\downarrow j}_{tell}  A \;\mathit{\mathit{find\_Nemo}}.\]
When $M^{\downarrow j}_{tell} A \;\mathit{\mathit{find\_Nemo}} \in
\mathit{SPEC}(seeker)$ and, due to (\ref{assumption8}) and
(\ref{assumption8a}), we have that $j$ (in $\mathit{SPEC}(j))$ will
eventually obtain the delegation from the $seeker$:
 \[\sometime M^{\downarrow \mathit{seeker}}_{do} I \;\mathit{\mathit{find\_Nemo}}.\]
\mfchange{Due to (\ref{assumption9}), (\ref{assumption10}) and}
 (\ref{asumption7}) the $seeker$ will eventually be told (by $j$) that
 $I \;\mathit{\mathit{find\_Nemo}}$:
\[ \sometime M^{\downarrow \mbox{j}}_{tell} B \;\mathit{\mathit{find\_Nemo}}.\] 
Lastly, due to (\ref{assumption8}) and (\ref{assumption8a}), we obtain that $\sometime B \;\mathit{\mathit{find\_Nemo}} \in \mathit{SPEC}(seeker)$ \mfchange{and so, eventually, the $seeker$ will believe that Nemo has been found.}

\subsection{Software Analysis}
As a second example we consider a larger task\mfchange{, namely} checking
the correctness of a substantial piece of software. %% ; namely it does
%% what it is supposed to do and is bug free. 
The task is broken down into small \mfchange{``human intelligence''}
tasks of similar complexity each of which is \mfchange{then}
crowdsourced. Each software fragment is considered (in)correct if two
out of three crowd members agree on its classification. We need to
verify not only that the crowd will check the whole \mfchange{of the}
software, but also that the software can be kept confidential, namely
no individual crowd member can gain access to the full software, nor
deduce the whole \mfchange{of the} software from the fragments he or
she is checking.
 
\begin{figure}[t!]
        \centering
       \includegraphics[width=0.59\textwidth]{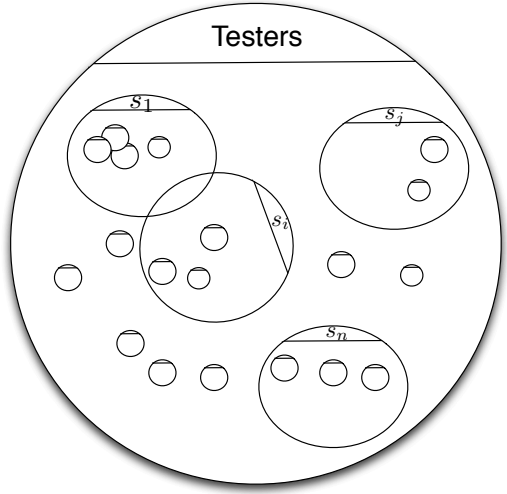}
                \caption{A crowd of specification testers.}
                \label{fig:testers}
        \end{figure}

Assume that the software $S$ is fragmented into $n$ chunks $\sigma_1,
\sigma_2, \ldots, \sigma_n$. To be able to participate in the process
an agent must have the required software skill level.  This is tested
as a condition of entry into the crowd. The crowd that tests the code
contains all the agents that have been so vetted. The structure of
this crowd, encapsulated in \mbox{Testers}, is given in
Fig.~\ref{fig:testers}. In the content of \mbox{Testers}, there are as
many context-agents as there are subsystems. Individual crowding
agents can choose which context they want to join, but they may not be
allowed to join some or all of them, based on other contents they are
members of. The $s_k$ agents send their content lists to
\mbox{Testers}.  Then, any content-respective software fragment is
only sent to members of the content.

The relations among the software fragments are given in the
specification of \mbox{Testers}, in form of $ (B \sigma_x\wedge \cdots
\wedge B \sigma_z) \rightarrow B\; whole$, for some $\{\sigma_x, \ldots, \sigma_z\} \subset \{\sigma_1,
\sigma_2, \ldots, \sigma_n\}$.  In addition, formulas
(\ref{tester1a}-\ref{tester2a}) are included as part of the
specification for \mbox{Testers}.
\begin{enumerate}
\item[$\ast$] Forward any incentives to the testers: 
  \begin{equation}\label{tester1a}
  M^{\downarrow s_i}_{adv} (A \;\mathit{test} \rightarrow \sometime A\; { earn })  \rightarrow  M^{\uparrow CN}_{tell} M^{\downarrow s_i}_{adv} (A \;\mathit{test} \rightarrow \sometime A\; { earn }).    
  \end{equation}

\item[$\ast$] $S$ is tested when each fragment is reported as tested:
  \begin{equation}\label{tester1b}
B\; tested(S) \leftrightarrow \big(  M^{\downarrow s_x}_{tell}  B\; tested(\sigma_x)   \wedge \cdots \wedge M^{\downarrow s_x}_{tell}  B\; tested(\sigma_z) \big). 
  \end{equation}

\item[$\ast$] If an agent is safe for a fragment $x$, it can be added
  to context $s_x$:
  \begin{equation}\label{tester1c}
 M^{\downarrow s_x}_{tell}  M^{\downarrow  i}_{tell} I\; cx(s_x) \wedge B\;  \mathit{safe}(i, s_x) \rightarrow M^{\uparrow s_x}_{tell} B\; ok(i). 
  \end{equation}

\item[$\ast$] An agent is safe for a fragment $x$, if it cannot deduce
  the whole software system from the fragment $x$ and the fragments
  already sent to it\footnote{If $B \; whole$ somehow becomes true independent of the specification provided by the testers then this formula will prevent any agent being deduced safe for any fragment.  This represents a ``fail safe'' situation -- no more fragments will be assigned and risk compromising the confidentiality of the software but obviously the task can not be completed in this circumstance.}:
  \begin{equation}\label{tester2a}
\neg \big( \big( \big( \bigwedge_{B\; \mathit{safe} (i, s_z) ,z\neq x } B \sigma_z \big) \wedge B \sigma_x \rightarrow B\; whole  \big)\big)  \rightarrow B\;  \mathit{safe}(i, s_x)  .
  \end{equation}
\end{enumerate}
The formulas (\ref{fragmentA}--\ref{fragmentF}) are  part of the
specification for $s_x$.
\begin{enumerate}
\item[$\ast$] An incentive is sent:
   \begin{equation}\label{fragmentA}
   M^{\uparrow Testers}_{adv} (A \;\mathit{test} \rightarrow \sometime A\; { earn }).    
   \end{equation}
 
\item[$\ast$] If an agent is interested in joining the context $s_x$,
  and the \mfchange{software} fragment is not yet tested, then \mbox{Testers} is told
  about this (and implicitly asked to approve the safety of the
  agent):
  \begin{equation}\label{fragmentB}
\big(M^{\downarrow  i}_{tell} G\; cx(s_x) \wedge \neg B\; approve(\sigma_x) \wedge \neg B\; \mathit{reject}(\sigma_x)\big) \rightarrow M^{\uparrow Testers}_{tell}  M^{\downarrow  i}_{tell} I\;  cx(s_x).
  \end{equation}

\item[$\ast$] If an agent is vetted, it is entrusted with a fragment
  for testing:
  \begin{equation}\label{fragmentC}
 M^{\downarrow s_j}_{tell} B\; ok(i) \rightarrow M^{\uparrow i}_{do} I\; test(\sigma_x).
  \end{equation}

\item[$\ast$] If two out of three agents 
  \mfchange{($i$, $j$ and $k$ are different)} approve the fragment, then
  the fragment is considered as being approved:
  \begin{eqnarray}\label{fragmentD}
 \big(M^{\downarrow i}_{tell} B\; approve(\sigma_x)  \wedge   M^{\downarrow j}_{tell} B\; approve(\sigma_x) \wedge (M^{\downarrow k}_{tell}B\; approve(\sigma_x) \vee   \\
 M^{\downarrow k}_{tell}B\;\mathit{reject}(\sigma_x))\big) \rightarrow  B\; approve(\sigma_x). \nonumber 
  \end{eqnarray}
  
\item[$\ast$] If two out of three agents 
  \mfchange{(again, $i$, $j$ and $k$ are different)} 
  reject the fragment, then the
  fragment itself is rejected:
  \begin{eqnarray}\label{fragmentE}    
          \big(M^{\downarrow i}_{tell} B\; \mathit{reject}(\sigma_x)  \wedge   M^{\downarrow j}_{tell} B\; \mathit{reject}(\sigma_x) \wedge (M^{\downarrow k}_{tell}B\; approve(\sigma_x)  \vee \\
          M^{\downarrow k}_{tell}B\;\mathit{reject}(\sigma_x))\big)  \rightarrow  B\; \mathit{reject}(\sigma_x). \nonumber
  \end{eqnarray}
 
\item[$\ast$] When the whole code fragment is approved or rejected,
  \mbox{Testers} is informed:
  \begin{equation}\label{fragmentF}   
(B\; approve(\sigma_x) \vee B\; \mathit{reject}(\sigma_x))  \rightarrow M^{\uparrow Testers}_{tell}  B\; tested(\sigma_x).   
  \end{equation}
\end{enumerate}
Finally, every crowd-member is assumed to have the formulas
(\ref{crowderA}--\ref{crowderF}) in its specification:
\begin{enumerate}
\item[$\ast$] It is an agent's goal to earn, and it is able to test
  software:
  \begin{equation}\label{crowderA}
   G\; earn\ \wedge\ A\; test .
  \end{equation}

\item[$\ast$] If it is able to test software and it actively pursues the
  goal to test software then eventually the software will be either
  approved or rejected:
  \begin{equation}\label{crowderB}   
  A \; \mathit{test}(\sigma_x) \wedge I \; \mathit{test}(\sigma_x)  \rightarrow  \sometime  \big (B\; approve(\sigma_x) \vee     B\; \mathit{reject}(\sigma_x) 
\big).
  \end{equation} 

\item[$\ast$] If an incentive is received, then pursue this to seize
  the opportunity:
  \begin{equation}\label{crowderC}   
 M^{\downarrow Testers}_{tell} M^{\downarrow s_x}_{adv} (A \;\mathit{test} \rightarrow \sometime A\; { earn })  \rightarrow M^{\uparrow s_x}_{tell} G\;  cx(s_x).
  \end{equation} 

\item[$\ast$] If delegated a \mfchange{code} fragment to test, then adopt the
  intention to do so:
  \begin{equation}\label{crowderD}   
 M^{\downarrow s_x}_{do}  I \; test(\sigma_x)  \rightarrow I \; test(\sigma_x). 
  \end{equation}
 
\item[$\ast$] Any code fragment can be either approved or rejected,
  but not both:
  \begin{equation}\label{crowderE}      
( B\; approve(\sigma_x) \wedge  \neg B\; \mathit{reject}(\sigma_x)) \vee ( \neg B\; approve(\sigma_x) \wedge B\; \mathit{reject}(\sigma_x) ).
  \end{equation}

\item[$\ast$] When the code is tested, send the results to the
  relevant context:
  \begin{eqnarray}\label{crowderF}  
    B\; approve(\sigma_x) \rightarrow M^{\uparrow s_x}_{tell}     B\; approve(\sigma_x).  \\
    B\; reject (\sigma_x) \rightarrow M^{\uparrow s_x}_{tell}     B\; \mathit{reject}(\sigma_x).   \nonumber
\end{eqnarray}
\end{enumerate}
Establishing that (\ref{tester3}) holds for \mbox{Testers} 
\mfchange{should then be}
straightforward.
\begin{equation}\label{tester3}
\sometime   B\; tested(S) \wedge \always  \big( M^{\uparrow s_x}_{tell} B\; ok(i)  \rightarrow B\;  \mathit{safe}(i, s_x) \big).  \\
\end{equation}
%
%  \section{Related work}\label{related}
% 
%     %wait for ijcai papers.
%     
%     verification of swarms
%     
%     dynamic epistemic logic
%     

\subsection{Utilising the Formal Basis}
Above we have given two examples showing both how our formal syntax
can be used to specify digital crowd scenarios, and then how formal
verification processes can use these specifications to establish
properties. The important aspects concerning this process are:
\begin{enumerate}
\item the general properties of the crowd, and environment, are
  formalised using our language;
\item the specific properties of individual agents within the digital
  crowd are also formalised using the same language;
\item we also formalise any assumptions we make about agent (hence,
  human) or environmental behaviour;
\item for an individual agent, formal verification of its properties
  can be carried out by invoking existing agent model-checking tools
  such as~\cite{MCAPL_journal}, which prove that all possible
  executions of the agent conform to the logical requirements;
\item once proved for individual agents, reasoning about combined
  properties can take place, either using manual proof (as above) or
  by some form of automated process (note there are many automated
  provers for modal and temporal logics); and
\item once complete, we have verified properties of the digital crowd
  scenario under the assumptions specified and so can be sure what
  behaviour will occur (if the assumptions are satisfied).
\end{enumerate}
Once this process is complete, it is natural to then revisit the
assumptions, weaken them and see if the verification can still be
carried through. If it cannot, then we can see where the required
behaviour can fail and so can either take this back to the crowd
application designer, or weaken the properties being verified. And so
on. The process continues in this way until sufficient verified
behaviour has been extracted. Note that we clearly cannot completely
specify all crowd behaviour, but can explore classes of behaviour and
prove what will occur under these assumptions.

\section{Summary}\label{summary}
Over the last few years, the business practice of crowdsourcing has
begun to capture the attention of computer scientists. All
crowdsourcing processes involve the participation of a digital crowd,
a large number of agents that access a single Internet platform or web
service. Although a digital crowd is a collection of agents, it is not
a structure traditionally studied within the area of multi-agent
systems. Logic-based formal methods are available for analysing the
behavior of, and the dynamics within, a multi-agent system before the
actual system is constructed. In particular, \mfchange{the formal}
verification of systems is the process of establishing that a designed
system has the intended properties. Crowdsourcing can be made more
reliable and effective by applying such logic-based formal methods by,
for example, determining important properties of a digital crowd,
under given assumptions about its members, before the crowd is
assembled.
  
Our aim is to enable automated formal verification, in particular
model-checking, to be utilised in digital crowd systems, and
crowdsourcing in general. To this end we extend the paradigm of an
agent, in particular a $\mathit{BDI}$ agent, as used in traditional
multi-agent systems.  Our extended agent encompasses both
communication behaviour and further individual agents and
sub-structures. We accomplish this by abstracting from the member
agents' mental states and directly specifying the communication
exchanged by the agents.
 
We proposed an \mschange{abstract} agent specification language, adding the representation
of messages to a temporal $\mathit{BDI}$ language.  Although most
agent programming languages allow messages to be passed among agents,
these messages are primarily signals (to \emph{stop}, \emph{go}, etc)
and are not part of the reasoning within the agent. \mschange{  The content of agent communication typically is not considered a separate informational state of the agent. We}
represent the messages in the same fashion as the informational,
motivational and dynamic mental attitudes of the $\mathit{BDI}$ agent
are represented. While nothing is assumed about the mental states of a
particular agent, the information it exchanges with other agents can
be sufficient to reason about a structure in which that agent belongs,
and can represent loose social structures such as digital crowds. To
exhibit this, we develop two examples to illustrate how our formal
specification language can be used as the basis for 
\mfchange{formally describing}
properties of a digital crowd.
  
The possibility of applying formal methods to digital crowds opens
many interesting avenues for future research.
%% We hope to use model-checking since it is an automated technique with good tool support.  
Model-checking requires the presence of an executable model of the
system under investigation.
%% In order to provide such a model we will need to develop a theory for modelling the membership of a crowd (\eg how many individual agents are needed in the model to adequately capture the behaviour of a crowd of arbitrary size).  
There are executable specification languages for agent systems based
on temporal logic~\cite{Fisher:1994,Shoham:1993} and we are
interested in adapting these so that our specifications can be easily
converted into appropriate executable models.  We anticipate that we
will need to develop abstraction mechanisms \eg a suitable crowd size
for a model may be the number of distinct crowd members referred to in
the property description. Extending the problem with probabilistic
aspects would also be of interest. This would enable reasoning about
the probable success of some task given to a digital crowd based on
(stochastic) assumptions about the probable abilities (and
reliability) of members of the crowd.  Probabilistic Model-checking
tools (\eg {\sc Prism}~\cite{kwiatkowska02prism}) could be adapted to
study digital crowds.

Finally, while we do not need to axiomatize our specification language
to use it in model-checking it would be interesting to develop an
axiomatization for the $M$ operators and formally study its
properties. We currently make no use of plans and \mfchange{planning} in our
examples, though planning is an important part of multi-agent systems
and it would be interesting to integrate the problem of planning with,
and for, digital crowds.

 \bibliographystyle{plain}
\bibliography{references}
\end{document}